\documentclass[12pt]{article}
\usepackage{amsmath}
\usepackage{amssymb}
\usepackage{epsfig}
\usepackage{cite}
\usepackage{color,colordvi}

\begin{document}
\vspace{0.01cm}
\begin{center}
{\Large\bf Landau-Ginzburg Limit of Black Hole's Quantum Portrait: Self Similarity and Critical Exponent} 

\end{center}

\vspace{0.1cm}

\begin{center}

{\bf Gia Dvali}$^{a,b,c,d}$\footnote{georgi.dvali@cern.ch},and {\bf Cesar Gomez}$^{a,e}$\footnote{cesar.gomez@uam.es}

\vspace{.6truecm}


{\em $^a$Arnold Sommerfeld Center for Theoretical Physics\\
Department f\"ur Physik, Ludwig-Maximilians-Universit\"at M\"unchen\\
Theresienstr.~37, 80333 M\"unchen, Germany}


{\em $^b$Max-Planck-Institut f\"ur Physik\\
F\"ohringer Ring 6, 80805 M\"unchen, Germany}

{\em $^c$CERN,
Theory Department\\
1211 Geneva 23, Switzerland}


{\em $^d$Center for Cosmology and Particle Physics\\
Department of Physics, New York University\\
4 Washington Place, New York, NY 10003, USA}

{\em $^e$
Instituto de F\'{\i}sica Te\'orica UAM-CSIC, C-XVI \\
Universidad Aut\'onoma de Madrid,
Cantoblanco, 28049 Madrid, Spain}\\

\end{center}


\begin{abstract}
\noindent  
 
{\small
 Recently we have suggested that the microscopic  quantum description of a  black hole is  an overpacked  self-sustained Bose-condensate of 
$N$ weakly-interacting soft gravitons,  
which obeys  the rules of 
't Hooft's  large-$N$ physics. 
 In this note we derive an effective Landau-Ginzburg Lagrangian for the condensate and 
 show that it becomes an exact description in a semi-classical  limit that serves as  the black hole analog of 't Hooft's planar limit.  The role of a weakly-coupled  Landau-Ginzburg order parameter is played by $N$.  This description consistently  reproduces the known properties of black holes in semi-classical  limit.  
Hawking radiation, as the quantum depletion of the condensate, is described  by the slow-roll of the field $N$. In the semiclassical limit, where black holes of arbitrarily small size are allowed, the equation of depletion is self similar leading to a scaling law for the black hole size with critical exponent $\frac{1}{3}$. }

\end{abstract}

\thispagestyle{empty}
\clearpage



We have recently composed \cite{gia-cesar} a quantum portrait of a black hole as a Bose-condensate of 
soft weakly-coupled gravitons.   The condensate is characterized by a typical wave-length which can assume different values.  Nevertheless what makes
it very special is that the occupation number of gravitons, $N$,  is always the maximal possible compatible with a given wave-length.  Putting it simply, the black hole is 
the maximally packed system in nature.  This overpacking makes the quantum theory of the black hole extremely simple.   The black hole is a large-$N$ system in 't Hooft's sense \cite{tHooft}. 
 To make a concrete analogy, the  quantum nature of black holes  is very similar to Witten's picture of  baryons  in $SU(N)$ gauge theories\cite{Witten}.  The crucial difference is that in a gauge theory $N$ is an input parameter, but for black holes it can take an arbitrarily large value. 
 The reason is that the  graviton coupling depends on its wave-length and therefore can be self adjusted in order to balance the kinetic energy of an  individual graviton  versus  the 
 collective binding potential  of $N$ gravitons resulting into a self-sustained condensate 
 for arbitrary $N$.  This self-sustainability condition requires fulfillment  of certain relations among the parameters of the condensate  that  turn out to be exactly of the same kind as those defining large-$N$ in 't 'Hooft's sense.
  
   This is a huge bonus which gravity provides us as compared to gauge theories. 
 To make  this advantage obvious,  let us note that the quantum portrait of  a black hole of the earth's mass  is as simple as of the baryon in  $SU(10^{66})$ gauge theory!.
 
   Defining the Planck length ($L_P$) and the corresponding Planck mass  ($M_P$) via Newton's coupling constant $G_N$ as, 
   \begin{equation}
   L_P \, \equiv \, {\hbar \over M_P} \,  \equiv \, \sqrt{ \hbar \, G_N }\, , 
   \label{planck}
   \end{equation}
  we can list all the characteristics of the black hole in terms of $L_P$ and $N$ as follows, 
     
     \begin{itemize} 
   \item    Occupation number  $=\,  N$ 
   \item    Wave-length $= \,  \sqrt{N} L_P$
  \item  Coupling stength $=  \,  1/ N$
  \item  Mass   $= \,  \sqrt{N}  {\hbar \over L_P} $
  \item Temperature $ = \, {\hbar \over \sqrt {N} L_P} $
 \item Entropy  $= \, N$
  \end{itemize}

 The occupation number can be written in  several different equivalent ways as \cite{class1, gia-cesar}
 \begin{equation}
  N \, = \, M^2 G_N/\hbar \, = \, M^2/M_P^2  \, .
  \label{Number}
  \end{equation}
  
  We have shown that this picture gives a fully quantum description of the known properties of black holes  that have been previously obtained only in the semi-classical limit (see below). 
  For example,  Hawking radiation with its characteristic negative heat capacity can be understood 
  in very simple terms as the quantum depletion of the Bose-condensate. The products of the depletion imitate the thermal spectrum despite the fact that the condensate itself  is cold. 
  We are not going to repeat the discussion about the quantum portrait and refer the reader 
  to the original paper \cite{gia-cesar} for the detailed discussion of all the aspects.  
  
  The question we want to address in this note  is the following. If a black hole is a Bose-condensate 
  it must admit a corresponding Landau-Ginzburg description. 
   What is this description? 
  
   We shall now display this theory.   First,  we need to identify the correct Landau-Ginzburg order parameter.  We start by making an intelligent guess that this must be $N$. 
   This guess turns out to be correct. 
  We need to obtain an effective Landau-Ginzburg Lagrangian that would correctly account for the (corse-grained) dynamics of $N$.   The key relies on the {\it leakiness} of the condensate i.e $N$ 
  constantly decreases due to the quantum depletion.  In the language of Landau-Ginzburg theory 
  this depletion should be translated as a slow-roll of the order parameter $N$ down the 
  slope of  an effective 
  potential. 
   In order to reconstruct this potential, we use the quantum depletion law derived in \cite{gia-cesar}. This law reads, 
   \begin{equation}
   {\dot N} \, = -  {1 \over \sqrt{N} L_P}  \, + \, {\mathcal O} (N^{-3/2}) \, , 
   \label{deplete}
   \end{equation}
 where dot stands for the time-derivative. 
 The leading term in this depletion comes from the two graviton scattering process, which in 
 a certain sense corresponds to a gravity equivalent of 't Hooft's planar limit. This process is a quantum progenitor of what in the semi-classical (planar)  limit 
 becomes the famous Hawking radiation.  
 
 The key to reconstruct the Landau-Ginzburg theory is to promote $N$ into a field 
 and then read-off the Lagrangian from the depletion equation (\ref{deplete}). This is very easy, and the corresponding Lagrangian has the following simple form, 
 \begin{equation}
  {\mathcal L}_{LG} \, = \,  (\dot{N})^2  \, + \,  {1 \over N} L_P^{-2}  \, +  \, L_P^{-2} {\mathcal  O}(1/N^2) \, .
 \label{LG}
 \end{equation} 
  The remarkable thing about this Lagrangian is that it fully exploits the power of large-$N$ physics.  
 Notice, that  since the only dimensionfull parameter is  $L_P$,  all the non-derivative terms 
 are multiplied by $L_P^{-2}$.    As a result all higher powers of  $\frac{1}{N}$ vanish in the 
 semi-classical limit in which $N \, \rightarrow \, \infty $, but a quantity $R \, \equiv \, \sqrt{N}\, L_P $ is kept fixed. 
Thus the semi-classical limit is analogous  to 't Hooft's  planar limit  with  $R$ playing a role similar to the
QCD length-scale that is set by t'Hooft's coupling. 

  To see that in the above "planar" limit  our Landau-Ginzburg  Lagrangian indeed recovers the 
  well-known semi-classical black hole physics, let us take the limit carefully. 
   We thus  take, 
   \begin{equation}  
   N \, \rightarrow \, \infty, ~~~L_P\, \rightarrow \, 0\,,   ~~~ R\equiv \sqrt{N}L_P \, =\, {\rm finite}, ~~~\hbar \, = \, {\rm finite}\, .
   \label{limit}
   \end{equation}
  In this limit the  Landau-Ginzburg Lagrangian acquires the following  exact form, 
   \begin{equation}
  {\mathcal L}_{LG} \, = \,  (\dot{N})^2  \, + \,  {1 \over N} L_P^{-2}  \,.  
 \label{LGexact}
 \end{equation} 
 Notice, that the scale $R$ was introduced simply as a parameter analogous to the QCD scale 
 that is kept fixed in  't Hooft's planar limit. 
 However, in the semiclassical limit it acquires  a geometric meaning.  To see this, let us rewrite 
 the depletion law (\ref{deplete}) in this limit, 
   \begin{equation}
   {\dot N} \, = -  {1 \over  R}  \, . 
   \label{depleteexact}
   \end{equation}
 Rewriting now $N$ in terms of the black hole mass using (\ref{Number}) we get 
    \begin{equation}
   {\dot M} \, = -  {\hbar  \over  R^2}  \, . 
   \label{depletemass}
   \end{equation}
 The latter expression is nothing but the Hawking's evaporation law in which 
 $R$ plays the role of the Schwarzschild radius,  and  $T \, =  \, {\hbar  \over R}$ is the 
 temperature. 
 
 A very interesting property of the depletion equation in the semiclassical limit is that it is self similar with respect to the scaling transformations:
 \begin{equation}
 N\rightarrow \lambda^{\frac{2}{3}} N  \, ~~~
 {\rm and} \, ~~~  t \rightarrow \lambda t \, .
 \end{equation}
 The deep root of this self-similarity lies in the quantum properties of the portrait. Namely, since we can have black holes with arbitrary value of $N$ and all of them are characterized by just this single parameter the process of depletion must be {\it self-similar}  with respect to changes of $N$ at least for large  $N$. This self-similarity becomes an exact property in the planar limit (\ref{limit}). 
 Changing to an energy variable $E \, \equiv \, \frac{\hbar}{t}$ we can rewrite the depletion equation as a renormalization group flow of $N$ in energy, namely
 \begin{equation}
 E^2 \frac{dN}{dE} \, = \,   \frac{M_P}{\sqrt{N}} \, .
 \end{equation}
 The solution to this equation is (the coefficient  $3/2$ is dropped)
 \begin{equation}
 N(E) \, = \,  \left (C - \frac{M_P}{E} \right )^{\frac{2}{3}} \, , 
 \end{equation}
 with $C$ an integration constant. Note that $N$ is growing with $E$ starting at the critical value $E_{cr}=\frac{M_P}{C}$ where $N(E_{cr})=0$. We can think of $E_{cr}$ as the threshold for black hole formation. Therefore we can write the following scaling law
 \begin{equation}
 N(E)\,  = \, [( E_{cr}^{-1} -E^{-1})M_P]^{\frac{2}{3}} \, .
 \label{scalingN}
 \end{equation}
  This scaling law exhibits a striking similarity with well-known scaling laws in black hole  formation. 
 To make the contact clear,  let us rewrite the above equation in terms of parameter $R$
 and also introduce a notation $p \, \equiv\,  \hbar^3 / M_P^2E$. We get,  
 \begin{equation}
  R\, = \, (p_{cr} \, - \, p)^{1/3} \, . 
  \label{scalingR}
 \end{equation}
 The above scaling is very similar to the  scaling law for black hole formation in classical general relativity  which was discovered by Choptuik \cite{choptuik} in the special case of  a spherical collapse. This result has been further generalized to many different types of gravitational collapse. The critical Choptuik exponent is  defined by $R\, = \, (p_{cr} - p)^{\gamma}$ with $p$ some control parameter and $p_{cr}$ the critical value. The value of the exponent is with small variations, of the order $\gamma = 0.36$. Comparing with (\ref{scalingR}) we get as the Landau-Ginzburg  prediction for the critical exponent $\gamma = \frac{1}{3}$.
 \footnote{Incidentally the Landau-Ginzburg exponent can be easily generalized to $4+d$ dimensions with the result $\gamma=\frac{1}{3+d}$. In $5$ dimensions this is in agreement with the result in \cite{Luis}.}
 
 It is important to stress here that the scaling law as well as the underlying self similarity are properties that are only exact in the Landau-Ginzburg  planar  (semi-classical)  limit  (\ref{limit}). In particular this makes at once clear  that the singularity  problem is an artifact of this  semiclassical limit.  Indeed only in this limit the evaporation law  (\ref{deplete}) is exact and says  that black holes of arbitrarily small $R$ can radiate, 
   leading to infinite temperature at $R=0$.  
 It is only in this planar limit that an idealized geometric concept, such as horizon,  becomes exact. 
 

    The effective description in terms of a weakly-coupled Landau-Ginzburg theory is also an important step 
    in closing the circle that connects the black hole's large-$N$ portrait with the idea of self-completion \cite{self} by classicalization \cite{class}.   The key concept is that the deep-UV theory 
    replaces the would-be strongly coupled degrees of freedom, with the  collective ultra-soft weakly coupled modes of $N$-particle states \cite{class1}.  The role of such a mode is played by the above  Landau-Ginzburg field $N$. 
 The notion of the ordinary Wilsonian renormalization flow cannot be defined  above  $M_P$ due to the absence of weakly-coupled one-particle states of Compton wavelength $\ll \,  L_P$.   This concept gets  replaced by a "running"  collective parameter  $N$.  In a very crude sense the $1/N$ dependence of the effective Landau-Ginzburg potential can be regarded as some sort of UV-fixed point in this non-Wilsonian "RG"-flow. 
 
 In summary the quantum $N$-portrait of black holes provides the quantum basis for a description of gravity  in terms of entity  that is analogous to  a master field \cite{master}. In the semiclassical (planar)  limit  (\ref{limit}) this master field is governed by the Landau-Ginzburg  Lagrangian (\ref{LG}). This naturally leads to self similarity and to black hole scaling laws.
 We thus observe that these scaling laws may be  rooted in large-$N$ nature of quantum black hole physics.

\section*{Acknowledgements}

It is a pleasure to thank Luis Alvarez-Gaume  and Slava Mukhanov  for discussions.
The work of G.D. was supported in part by Humboldt Foundation under Alexander von Humboldt Professorship,  by European Commission  under 
the ERC advanced grant 226371,   by TRR 33 \textquotedblleft The Dark
Universe\textquotedblright\   and  by the NSF grant PHY-0758032. 
The work of C.G. was supported in part by Humboldt Foundation and by Grants: FPA 2009-07908, CPAN (CSD2007-00042) and HEPHACOS P-ESP00346.


\begin{thebibliography}{99}
    
 \bibitem{gia-cesar}    
 	
G.~Dvali, C.~Gomez,  Black Hole's Quantum N-Portrait, 
 arXiv:1112.3359 [hep-th].  
    
    
 \bibitem{tHooft} G.~'t Hooft, A Planar Diagram Theory for Strong Interactions.
 Nucl.Phys. B72 (1974) 461
   
    
  \bibitem{Witten}
E.~ Witten, 	
Baryons in the 1/n Expansion. Nucl.Phys. B160 (1979) 57

\bibitem{class1}

G.~ Dvali, C.~ Gomez, Alex Kehagias.Classicalization of Gravitons and Goldstones.
arXiv:1103.5963 [hep-th],  JHEP 1111 (2011) 070



\bibitem{choptuik}
M. W. Choptuik, Universality And Scaling In Gravitational Collapse Of A Massless Scalar Field, Phys. Rev. Lett. 70 (1993) 9.

 
C. Gundlach, Critical phenomena in gravitational collapse, Phys. Rept. 376 (2003) 339 [arXiv:gr-qc/0210101]. 

\bibitem{Luis}
L.~Alvarez-Gaume , C.~Gomez , A.~Sabio Vera, A.~Tavanfar , M.~Vazquez-Mozo,	
Critical formation of trapped surfaces in the collision of gravitational shock waves, JHEP 0902 (2009) 009 ,arXiv:0811.3969 [hep-th]

\bibitem{self}
G.~ Dvali  and C.~ Gomez, Self-Completeness of Einstein Gravity,  arXiv:1005.3497 [hep-th]


\bibitem{class}

G.~ Dvali, G.~ F. Giudice, C.~ Gomez, A.~ Kehagias,  UV-Completion by Classicalization, arXiv:1010.1415 [hep-ph]. JHEP 2011 (2011) 108 


\bibitem{master}
E. Witten, in Recent Developments in Gauge Theories eds. G. 't Hooft et. al. Plenum Press, New York and London (1980).

   
\end{thebibliography}
\end{document}